\documentclass[aps,prl,reprint,superscriptaddress,graphicx]{revtex4-1}% for checking your page length
\usepackage{amsmath}
\usepackage{graphicx}
\usepackage{color}
\usepackage{float}
\usepackage[separate-uncertainty=true,multi-part-units=single]{siunitx}
\newcommand*\mean[1]{\bar{#1}}

\draft % marks overfull lines with a black rule on the right

\definecolor{darkgreen}{rgb}{0.0,0.4,0.0}
\DeclareSIUnit{\mu}{\micro\meter}
\DeclareSIUnit{\unit}{\relax}
\begin{document}

% Use the \preprint command to place your local institutional report number 
% on the title page in preprint mode.
% Multiple \preprint commands are allowed.
%\preprint{}

\title{Fabrication of mirror templates in silica with micron-sized radii of curvature}

% repeat the \author .. \affiliation etc. as needed
% \email, \thanks, \homepage, \altaffiliation all apply to the current author.
% Explanatory text should go in the []'s, 
% actual e-mail address or url should go in the {}'s for \email and \homepage.
% Please use the appropriate macro for the type of information

% \affiliation command applies to all authors since the last \affiliation command. 
% The \affiliation command should follow the other information.

\author{Daniel Najer}
\author{Martina Renggli}
\author{Daniel Riedel}
\author{Sebastian Starosielec}
\author{Richard J. Warburton}
\affiliation{Department of Physics, University of Basel, Klingelbergstrasse 82, Basel 4056, Switzerland}
%\email[]{Your Mail address}
%\homepage[]{Your web page}
%\thanks{}
%\altaffiliation{}

% Collaboration name, if desired (requires use of superscriptaddress option in \documentclass).
% \noaffiliation is required (may also be used with the \author command).
%\collaboration{}
%\noaffiliation

\date{December 7, 2016}

\begin{abstract}
% AIP Guidelines: One paragraph, <500 words
We present the fabrication of exceptionally small-radius concave microoptics on fused silica substrates using $\text{CO}_2$ laser ablation and subsequent reactive ion etching. The protocol yields on-axis near-Gaussian depressions with radius of curvature $\lesssim\SI{5}{\mu}$ at shallow depth and low surface roughness of $\SI{2}{\angstrom}$. This geometry is appealing for cavity quantum electrodynamics where small mode volumes and low scattering losses are desired. We study the optical performance of the structures within a tunable Fabry-P\'{e}rot type microcavity, demonstrate near-coating-limited loss rates ($\mathcal{F} = \num{25000}$) and small focal lengths consistent with their geometrical dimensions.
\end{abstract}

% insert suggested PACS numbers in braces on next line $\lambda \approx \SI{10}{\mu}$

\maketitle %\maketitle must follow title, authors, abstract and \pacs

%\section{Introduction}
%\cite{ReithmaierNature2004}
%\cite{YoshieNature2004}
%\cite{HennessyNature2007}
%\cite{HungerAIPADV2012}
%\cite{BarbourJAP2011}
%\cite{GreuterAPL2014}
%\cite{GreuterPhysRevB2015}
%\cite{TrichetOptExp2015}
%\cite{HouApplOpt2015}
%\cite{PanJMO2008}
%\cite{LeeNatComm2012}

The physics of single emitters strongly coupled to optical resonators offers a rich variety of quantum applications, including high-brightness indistinguishable single photon sources, single-photon transistors, and photon-mediated emitter-emitter coupling. A significant but challenging area is the application of these concepts, initially developed in the context of atomic physics, to solid-state systems such as quantum dots, nitrogen-vacancy centers in diamond or localization centers in 2D materials. The key requirements to achieve a high Purcell enhancement or a coherent exchange of energy quanta---the so-called weak and strong coupling regimes of cavity quantum electrodynamics, respectively---are a small mode volume ($V$, the effective extension of the confined electromagnetic field), a high resonator quality factor ($\mathcal{Q}$, the average number of coherent oscillations) as well as a precise in-situ tuning between emitter and resonator~\cite{VahalaNature2003}. Various high-$\mathcal{Q}$ microresonator concepts exists, e.g.\ micropillars~\cite{ReithmaierNature2004} or photonic crystal cavities~\cite{YoshieNature2004}. The tunable Fabry-P\'{e}rot type microcavity uniquely offers full spectral and spatial tuning to the emitter~\cite{Trupke2005,HungerNJP2010} as well as highly efficient external mode-matching in the free-beam version~\cite{BarbourJAP2011}. Already, significant emitter-cavity cooperativity has been achieved with scaled down dimensions~\cite{GreuterPhysRevB2015}. For these tunable plano-concave microcavities, the route to high coupling rates lies within the strong confinement of the resonant mode which is achieved by reducing the curved mirror's radius of curvature ($R$).

Only a few fabrication methods are known to produce concave microoptics with radii of curvature $R \le \SI{5}{\mu}$, e.g.\ focussed ion beam (FIB) milling~\cite{TrichetOptExp2015, Kelkar2015}, femtosecond laser wet etching~\cite{ChenOptLett2014} and proximity-effect-assisted reflow techniques~\cite{HeOptLett2004}. For an optical cavity with total round-trip losses $L_\text{tot}$ and ultra-high finesse $\mathcal{F}\approx \pi/L_\text{tot}\approx 10^5$, acceptable round-trip scattering losses are below $S \approx (4 \pi \sigma / \lambda)^2 = \SI{30}{ppm}$, setting the upper limit for the surface roughness to a demanding $\sigma = \num{2}\ldots\SI{6}{\angstrom}$ in the VIS--NIR spectral range~\cite{NussbaumPureApplOptics1997}. The two latter approaches suffer from a too high surface roughness on the scale of a few nanometers. For FIB milling, microoptics fabrication benefits from full shape control and low surface roughness, and indeed depressions with curvatures as low as $R = \SI{1.5}{\um}$ have been reported recently~\cite{TrichetOptExp2015}. Yet FIB represents a production method at significant investment costs.

\begin{figure}
\includegraphics[width=85mm]{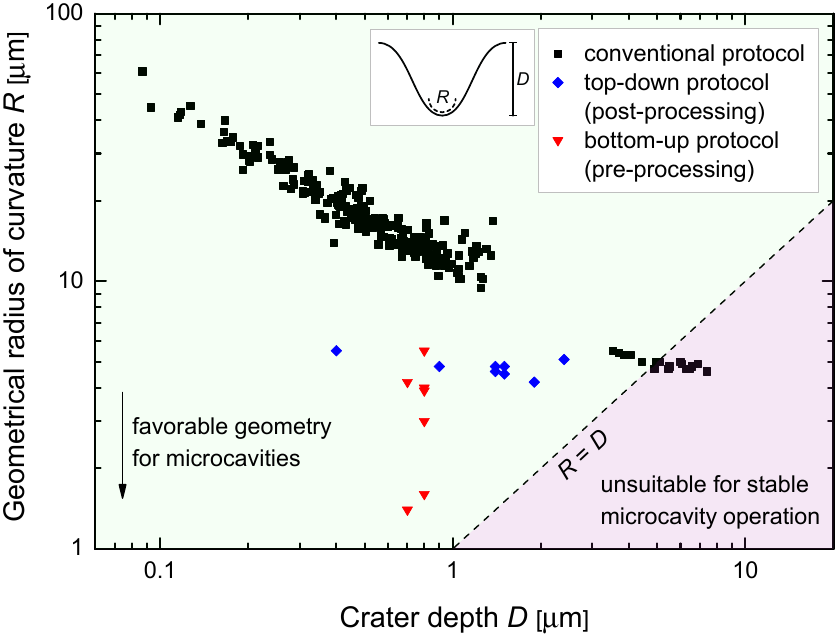}
\caption{Geometry parameters of craters produced by different fabrication methods. Conventional $\text{CO}_2$ laser ablation on polished fused silica substrates (black dots) shows a strong correlation between geometrical radius of curvature $R$ and depth $D$. Reproduced from the author's previous work, \citet{GreuterAPL2014}. In this work, the favourable regime of low $R$ at shallow $D$ becomes accessible by both post-processing (``top-down'', blue dots) and pre-processing (``bottom-up'', red dots) methods with a minimum radius of curvature $R = \SI{1.2 \pm 0.1}{\mu}$.}
\label{fig:fig1}
\end{figure}

$\text{CO}_2$ laser ablation on silica has turned out to be a low-cost alternative providing near-Gaussian shaped concave depressions at ultra-low surface roughness. In the conventional fabrication protocol, short pulses of $\text{CO}_2$ laser light are focussed onto either a polished fused silica substrate or the end facet of an optical fiber~\cite{HungerAIPADV2012}. At these wavelengths ($\lambda \approx \SI{10}{\mu}$), the light is efficiently absorbed by the material. While the dynamics of the process are complex, the following phenomenological description is well accepted: above a certain local temperature threshold, local evaporation results in a concave landscape roughly following the spatial irradiance distribution. The strong surface tension of the molten material then leads to a very smooth solidification process, with excellent surface roughness routinely at $\num{1}\ldots\SI{2}{\angstrom}$~\cite{HungerAIPADV2012,GreuterAPL2014}. 

The established $\text{CO}_2$ laser ablation process can produce a large range of depression geometries (black dots in figure~\ref{fig:fig1}), yet with a strong but unfortunate correlation between radius $R$ and depression depth $D$ (from \citet{GreuterAPL2014}) irrespective of varying fabrication parameters (intensity, pulse duration and pulse train length). While the radius seems to stagnate at $R \approx \SI{5}{\mu}$ or slightly below, a formed microcavity resonance becomes intrinsically unstable for cavity lengths $L > R$ (or $2R$) in a plano-concave (or biconcave) geometry from Gaussian optics theory. In practice, the finite lateral extent of the mirrors makes this transition between stable and unstable cavity modes less abrupt. In addition, deviations of the concave mirror surface geometry from the isophase plane of the cavity field can induce a mode-mixing between transversal modes~\cite{Benedikter2015}, acting as additional loss channels and further restricting the stability region to $L \lesssim R/2$. Those constraints further limit the established ablation protocol to geometries of $R \gg \SI{5}{\mu}$.

In this work we present a post-processing (``top-down") protocol which lifts this rigid link between radius and depth in the conventional ablation method. The key concept is to introduce an etch step post-ablation to reduce the crater depth, while preserving the high-grade geometry of the conventional ablation. In addition, we demonstrate first steps towards greater shape control at very small radii by pre-processing the substrates (a ``bottom-up" approach) with standard photolithography and subsequent $\text{CO}_2$ laser polishing. While optical polishing has been demonstrated~\cite{Calixto2005, Choi2015b} with convex features down to $\SI{9.4}{\mu}$ in radius~\cite{Choi2015a}, our bottom-up approach achieves much smaller radii, down to $\SI{1.2}{\mu}$. We characterize the structures formed by both protocols by confocal scanning microscopy and atomic force microscopy, and find radii of curvature $R\approx \SI{5}{\mu}$ at shallow depths $D \approx \SI{1}{\mu}$, as shown in figure~\ref{fig:fig1} (blue and red dots). We apply a reflective coating to our structures and record optical transmission spectra of the formed microcavities. We verify the geometrical radii and show high-$\mathcal{Q}$ performance of a microcavity fabricated by the top-down protocol.

\begin{figure}
\includegraphics[width=85mm]{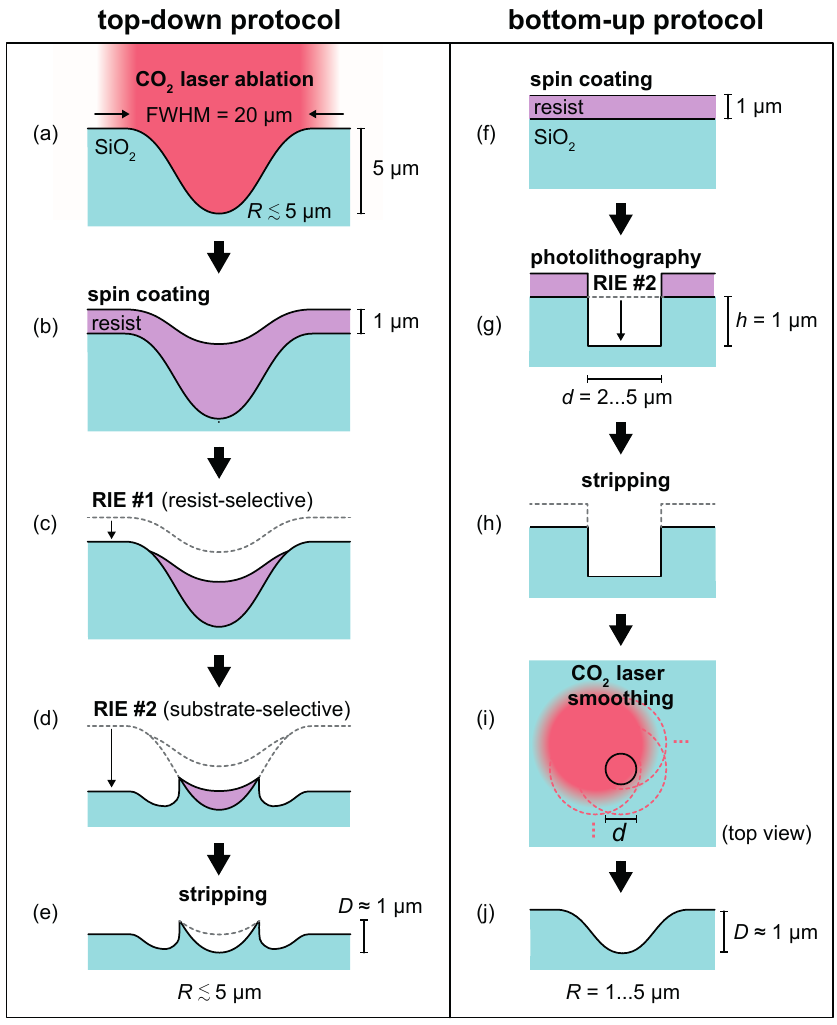}
\caption{Protocol for both processing schemes. The top-down approach starts from a small-radius crater at large depth (a) produced by conventional high-power $\text{CO}_2$ laser ablation with an FWHM beam diameter of $\SI{20}{\mu}$. With spin-coating of photoresist, the crater becomes partially filled (b). Reactive ion etching (RIE) with high selectivity for the resist consumes the resist homogeneously, implementing a self-centered mask for the crater (c). A second RIE step of low resist-selectivity attacks the exposed silica substrate, reducing the crater's depth (d). Stripping of the residual resist produces a crater at reduced depth, maintaining the original surface quality (e).
The bottom-up approach starts from a flat silica substrate, where a step-like crater template is formed by conventional photolithography (f-h). A coarsely scanned low-power $\text{CO}_2$ laser locally melts the silica surface (i) smoothing out the template shape into the desired form (j).}
\label{fig:fig2}
\end{figure}

%\section{Fabrication methods}
%\\subsection{Protocol 1: top-down approach}
%\label{sec:protocol1}
The top-down approach is based on the conventional ablation method in the regime of large depth and small radius ($R \approx \SI{5}{\mu}$) followed by a post-processing etching step to reduce the large depth by means of reactive ion etching (RIE), see figure~\ref{fig:fig2}(a--e).
After initial ablation with an FWHM beam diameter of $\SI{20}{\mu}$ (a), the depression is spin-coated with AZ1512HS photoresist (Microchemicals, Germany), which partially reflows into the depression (b). The first RIE etching step ($\text{Ar}$/$\text{O}_2$) has a preferred selectivity towards the photoresist, and after calibrated removal leaves a self-centered mask at the crater center (c). The second RIE step ($\text{CHF}_3$/$\text{Ar}$/$\text{O}_2$) preferentially etches the exposed silica substrate and thus reduces the effective depth of the ablation crater, while the crater center's ultralow-roughness surface and geometry are protected (d). A standard solvent stripping of the residual photoresist then reveals the original ablation crater bottom with radius $R \approx \SI{5}{\mu}$ at a reduced depth of $D \approx \SI{1}{\mu}$ with unaffected surface quality (e). Both RIE step durations are calibrated to the spin coating thickness, crater depth, and their corresponding etching rates.

%\\subsection{Protocol 2: bottom-up approach}
%\label{sec:protocol2}
The bottom-up approach, a first step towards shape control at $R\ll \SI{5}{\mu}$ using $\text{CO}_2$ laser ablation, relies on the optical polishing effect~\cite{NowakAO2006} on a pre-structured fused silica substrate. With low irradiation, ablation becomes negligible whereas surface-tension-induced smoothing remains effective~\cite{LaguartaApplOpt1994}. The protocol is sketched in figure~\ref{fig:fig2}. Photoresist (AZ1512HS) is spun onto a flat silica substrate (f) and locally removed by means of standard photolithography in order to expose the underlying silica for a substrate-selective RIE step ($\text{CHF}_3$/$\text{Ar}$/$\text{O}_2$) (g). After dissolving the residual resist, a binary structure with well-controlled width $d$ and height $h$ results (h). The pre-structured substrate is then optically polished by coarsely scanning the low intensity $\text{CO}_2$ laser irradiation (i) over the fabricated areas. Phenomenologically, the polished surface profile can be well described by a convolution of the template pattern with a Gaussian kernel of a remarkably small RMS width parameter ($s = \num{0.4}\ldots\SI{1}{\mu}$). For given $s$ and target radius $R$ we analytically estimate optimal template parameters ($d$,$h$) for a target paraboloid-like shape, i.e.\ we aim at creating a target depression shape with vanishing fourth order derivative at the crater center. For production values $s=\SI{1}{\mu}$ and target $R=\SI{3}{\mu}$ the best estimates of the feature dimensions are $d=\SI{3.5}{\mu}$ and $h=\SI{1.0}{\mu}$, a binary pattern easily accessible by standard photolithography.

Both protocols yield depressions with a small geometrical $R$ at shallow depth suited for a stable microcavity. For the main top-down protocol, the near-Gaussian shape, the low surface roughness as well as the high axial symmetry originating from the initial ablation process is well conserved. The prototype bottom-up protocol results in features with remarkably small radii $R = \num{1}\ldots \SI{5}{\mu}$ while exhibiting a reduced axial symmetry and a slightly increased surface roughness. We argue that these problems can be eliminated with a better smoothing procedure, as pointed out below.

%\section{Geometrical analysis}
\begin{figure}
\includegraphics[width=85mm]{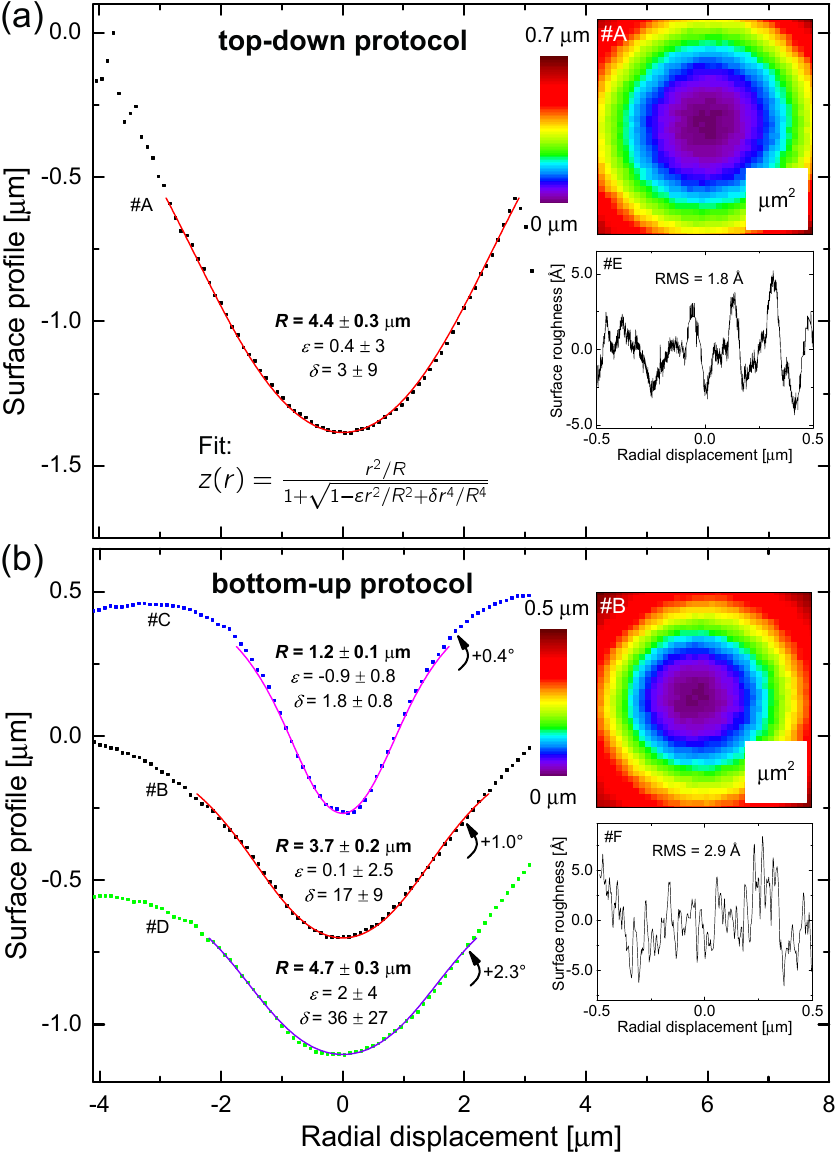}
\caption{Line scans (points) through top-down (a) and bottom-up (b) craters measured by confocal microscopy (blue and green curves shown with an offset). Axial asymmetry is compensated by introducing a small tilt to the data before fitting (see curved arrows). The solid lines depict cuts through the fits in which the surface is modelled by a modified ellipsoid $z(r)$. The uncercainty in the fitting parameters accounts for the variation on the fitting range around the center of the craters ($\num{3}\ldots \SI{6}{\mu}$). The insets show contour plots (non-tilted data) and AFM surface roughness measurements of selected craters revealing atomically smooth surfaces.}
\label{fig:fig3}
\end{figure}

We investigate the geometrical shape of the fabricated craters with confocal scanning microscopy and the surface roughness by atomic force microscopy. We describe the surface profile $z(r)$ as a function of radial displacement $r$ by means of a modified ellipsoid,
\begin{equation}
z(r) = \frac{r^2 / R}{1 + \sqrt{1 - \epsilon \, r^2 / R^2+\delta \, r^4 / R^4}} \;,
\label{eq.1}
\end{equation}
where $R$ represents the geometrical radius of curvature, and the dimensionless parameter $\epsilon$ is related to the conicity of the ellipsoid (e.g.\ $\epsilon=1$ for a sphere, and $\epsilon=0$ for a paraboloid). An ellipsoid is often used in the quantitative description of spherical and aspherical surfaces (e.g.\ ref.\ ~\citenum{NussbaumPureApplOptics1997} and~\citenum{ForbesOpticsExpress2007}), however this description becomes inappropriate at large $r$ when the optical surface converges into the unmachined flat substrate. We thus heuristically modify the ellipsoid description by another dimensionless parameter $\delta > 0$ to respect the surface asymptotics at $r \gg R$. For $r \ll R$ the ellipsoid is recovered.

Line scans through the center of craters \#A (fabricated by the top-down protocol), \#B, \#C and \#D (fabricated by the bottom-up protocol) are shown in figure~\ref{fig:fig3}. As seen from the area scan (insets), top-down crater \#A exhibits a high axial symmetry, while the axial symmetry of bottom-up crater \#B is slightly reduced. In order to fit the axially symmetric model to the bottom-up craters and extract the relevant geometric parameters, we apply a small tilt (up to $2.3{^\circ}$) to the profile data. As a result we retrieve $R^{\#A} = \SI{4.4 \pm 0.3}{\mu}$ for a top-down crater and $R^{\#B} = \SI{3.7 \pm 0.2}{\mu}$ for a bottom-up crater. The surface roughness measured by atomic force microscopy is \SI{1.8}{\angstrom} (top-down crater \#E~\cite{footnote}) and \SI{2.9}{\angstrom} (bottom-up crater \#F).

%\section{Optical analysis}
\begin{figure}
\includegraphics[width=85mm]{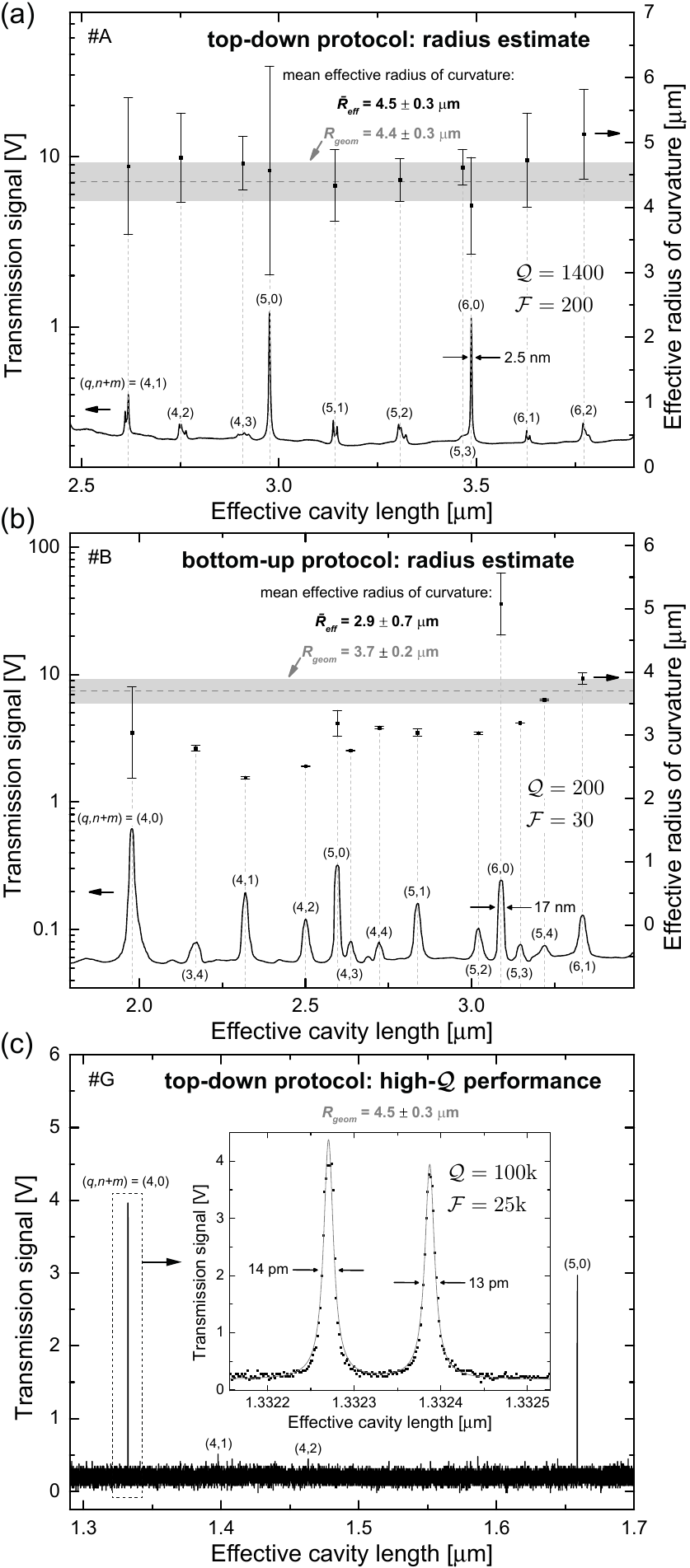}
\caption{Optical transmission spectra of tunable microcavities formed by fabricated top-down (a,c) and bottom-up (b) craters coated with $\text{Ti}$/$\text{Au}$ by electron beam evaporation (a,b), or a $\text{Ta}_2\text{O}_5/\text{SiO}_2$ quarter-wave stack by ion beam sputtering (c). The curved mirrors are paired with a polished silica substrate with the same coating. The mode-matching (a,b) is intentionally poor in order to couple the Gaussian input laser mode ($\lambda = \SI{940}{\nano\meter}$) to a large number of transversal cavity modes thus enabling the transversal mode splitting to reveal an effective radius of curvature $R_\text{eff}$. The high-$\mathcal{Q}$ configuration (c, $\lambda = \SI{637}{\nano\meter}$) shows a double-Lorentzian fine structure with a large quality factor $\mathcal{Q} = \num{e5}$ and high finesse $\mathcal{F} = \num{2.5e4}$.}
\label{fig:fig4}
\end{figure}

For optical characterization of the radius, the fabricated craters are coated with $\text{Ti}$(\SI{5}{\nano\meter})/$\text{Au}$(\SI{80}{\nano\meter}) by electron beam evaporation and paired with a polished silica substrate (with same coating) to form a tunable Fabry-P\'{e}rot type microcavity. The cavity transmission, with respect to length detuning induced by a piezo nanopositioner, is then measured~(see figure~\ref{fig:fig4}) at wavelength $\lambda = \SI{940}{\nano\meter}$. From Gaussian optics, $\text{TEM}_\text{qnm}$ resonator modes appear at cavity lengths
\begin{equation}
L_{qnm} = \left(q + \frac{n+m+1}{\pi} \cos^{-1}\sqrt{g} \right) \frac{\lambda}{2}\;,
\label{eq.2}
\end{equation}
where $q$ is the longitudinal mode index and $n,m$ are the transversal mode indexes. In the plano-concave cavity geometry, the confocal parameter is $g = 1 - L_{qnm}/R_\text{eff}$, which itself depends on the cavity length and allows for the extraction of an effective radius of curvature $R_\text{eff}$. 

As the mode overlap of the probing Gaussian laser beam to $\text{TEM}_\text{q00}$ is largest, an identification of the fundamental cavity modes ($n+m=0$) is straightforward. An intentionally poor mode-matching reveals the higher-order transversal cavity modes ($n+m\geq 1$) whose degeneracy is slightly lifted, indicating a small axial asymmetry of the crater. The (integer) longitudinal mode index $q$ is extracted from a wavelength detuning $\Delta\lambda$ ($\lambda = \num{918}\ldots \SI{974}{\nano\meter}$) and corresponding resonance shift $\Delta L$ via $q = \lfloor 2\Delta L/\Delta\lambda \rfloor$, where the non-integer residual of the right hand side originates from the $n+m+1$ contribution. The cavity length $L$ is calibrated from the experimental control parameter (the nanopositioner's piezo voltage) by extrapolation of the resonance modes to the (unphysical) limit $n+m+1 \rightarrow 0$ where $L_{qnm} \equiv q\lambda/2$. For each individual mode $L_{qnm}$ an effective radius $R_\text{eff}$ of curvature can be determined, with an uncertainty induced by the cavity length calibration procedure.

From cavity \#A (formed by the main top-down protocol), a mean effective radius of $\mean R^{\#A}_\text{eff} = \SI{4.5 \pm 0.3}{\mu}$ is found consistently across all probed cavity resonances (figure~\ref{fig:fig4}a), which closely matches the crater geometry $R^{\#A} = \SI{4.4 \pm 0.3}{\mu}$. The measured quality factor $\mathcal{Q} = 1400$ for the $\text{TEM}_\text{600}$ mode translates to a mirror reflectance of $\SI{98.6}{\percent}$ and a finesse of $\mathcal{F} = \lambda\mathcal{Q}/2L_\text{qnm} = \num{200}$, typical values expected for an evaporated $\text{Au}$ coating~\cite{OlmonPhysRevB2012}.

In contrast to these results, cavity \#B (formed by the prototype bottom-up protocol) shows a much broader range of effective radii with a mean of $\mean R^{\#B}_\text{eff} = \SI{2.9 \pm 0.7}{\mu}$ and significantly lower quality factors ($\mathcal{Q} = 200$ for the $\text{TEM}_\text{600}$ mode resulting in $\mathcal{F} = \num{30}$). While the mean effective radius lies significantly below the top-down threshold (\SI{5}{\mu}), which is also apparent from the geometrical analysis $R^{\#B} = \SI{3.7 \pm 0.2}{\mu}$, the strong fluctuation with cavity length indicates a deviation from a spherical geometry from the paraxial Gaussian beam theory. The reduction in the $\mathcal{Q}$ of the bottom-up cavity likely arises from an axial asymmetry and waviness which coincides with the coarse-scanned pitch (\SI{4}{\mu}) of the optical polish matrix. We expect this defect to be considerably lifted by a finer pitch in a refined fabrication run, the focus of further study. 

We test the high-$\mathcal{Q}$ performance of top-down crater \#G~\cite{footnote} with a commercial high-reflectivity quarter-wave stack $\text{Ta}_2\text{O}_5/\text{SiO}_2$ coating, produced by ion beam sputtering, at a wavelength of $\lambda = \SI{637}{\nano\meter}$. In figure~\ref{fig:fig4}(c), two adjacent longitudinal modes $(q,0)$ are recorded, each split into a linear-polarized fine structure likely originating from the slightly elliptical mirror surface~\cite{UphoffNJP2015}. For the $\text{TEM}_\text{400}$ mode, we demonstrate a large quality factor of $\mathcal{Q} = \num{e5}$ and high finesse $\mathcal{F} = \num{2.5e4}$, very well suitable for the target applications in cavity quantum electrodynamics. The measured finesse is close to the one expected from the bare mirror reflectivity ($\SI{99.9925}{\%}$, measured on a fused silica witness sample by the manufacturer).

%\section{Conclusion}
In conclusion, we presented two approaches to fabricate low-roughness, shallow, micrometer-sized concave mirror templates, thereby overcoming the present geometrical limits of conventional $\text{CO}_2$ laser ablation on polished fused silica substrates and on cleaved optical fibers. For the main method (top-down), we demonstrate high-finesse, stable microcavity operation at shallow depth and extract effective radii of curvature $\lesssim \SI{5}\mu$ consistent with their geometrical shape. The top-down protocol relies on the conventional ablation process and thus conserves its excellent geometrical properties such as a high axial symmetry and ultra-low surface roughness. The prototype bottom-up protocol demonstrates the early steps towards shape control at the few-micrometer level, presently unachieved by conventional $\text{CO}_2$ laser ablation alone. We speculate that the top-down protocol is robust enough for ultrahigh-finesse microcavity operation and see room for significant improvement in the bottom-up approach.

\begin{acknowledgments}
The authors acknowledge financial support from SNF (project 200020\_156637) and NCCR QSIT. 
\end{acknowledgments}

D.N. and M.R. contributed equally to this work.

%% Create the reference section using BibTeX:
%\bibliography{references}

\begin{thebibliography}{24}
\bibitem [{foo()}]{footnote}%
  \BibitemOpen
  \href@noop {} {\ }\bibinfo {note} {The tested top-down craters \#E and \#G
  belong to an earlier fabrication batch than \#A, where the photoresist was
  manually applied as a droplet. The resulting geometry, i.e.\ radius and
  depth, remains the same}\BibitemShut {NoStop}%
\end{thebibliography}
%% Run this once to generate your BBL file. Then copy the contents of your BBL file into your main latex file, commenting out "\bibliography"

%
\end{document}